# Acoustic jet based on acoustic metamaterial

**Liuxian Zhao[1], Timothy Horiuchi[1,2], Miao Yu[1,3,*]**

[1]Institute for Systems Research, University of Maryland, College Park, MD, 20742, USA

[2]Department of Electrical and Computer Engineering, University of Maryland, College Park, Maryland 20742, USA

[3]Department of Mechanical Engineering, University of Maryland, College Park, Maryland 20742, USA

lzhao128@umd.edu, timmer@umd.edu, mmyu@umd.edu

**ABSTRACT**

In this paper, a novel gradient index (GRIN) acoustic metamaterial is proposed based on the concept of optical modified generalized Luneburg lens (MGLL). With the MGLL, double-foci and high energy density between the two foci can be achieved, which enables the realization of an ultra-long acoustic jet between the two foci. Compared to the generalized Luneburg lens with a single focus, which can only achieve short acoustic jet in the near field or ultra-long acoustic jet in the far field, the proposed MGLL can achieve an acoustic jet extending from the near field to the far field. This capability of the MGLL is theoretically and numerically demonstrated. Furthermore, acoustic metamaterial based MGLL is designed with lattice unit cells having a variable filling ratio. Numerical simulation results show that ultra-long acoustic jets can be achieved with a jet length of up to $30\lambda$, covering both the near field and the far field.

## 1. Introduction

An acoustic jet is an acoustic focal field with a subwavelength full width at half maximum (FWHM) while maintaining a long propagation distance, which has potential applications in structural health monitoring [1] and medical imaging [2], and energy harvesting [3-5]. With the development of acoustic metamaterials for control and manipulation of the propagation of acoustic waves in recent years [6-9], using acoustic metamaterials to generate acoustic jet has attracted much attention. For example, Minin *et al*. theoretically proposed to use acoustic metamaterial lens to focus acoustic energy for achieving the acoustic jet [10]. Maznev *et al*. proposed to modify the aperture size of acoustic metamaterial lens to achieve acoustic jet, which was capable of focusing at least 50% of the incident power [11]. Lopes *et al*. experimentally demonstrated the use of an acoustic metamaterial lens for achieving an acoustic jet with an FWHM smaller than $\lambda/2$ ($\lambda$ is wavelength) [12]. Furthermore, Canle *et al*. proposed a practical realization of subwavelength acoustic jet by properly designing the metamaterial lens and choosing the lens and background medium materials [13]. Most existing methods for creating acoustic jets have a limitation on the working distance due to the small focal area. More recently, Lu *et al*. designed a gradient index (GRIN) acoustic metamaterial generalized Luneburg lens (GLL), which can achieve super long working distance up to $17\lambda$ [14]. However, this GRIN lens can only achieve super long acoustic jet in the far field or short acoustic jet in the near field.

Luneburg lens is a spherically symmetric gradient-index lens, which can focus incoming plane waves onto the outer surface of the opposite side of the lens [15-19]. Optical double-foci Luneburg lens was initially proposed in 1984 by Jacek Sochacki [20], which was derived from the standard optical Luneburg lens [21]. By tailoring the refractive index distribution, a modified generalized Luneburg lens (MGLL) can be designed to achieve two focal points at arbitrary positions, which allow high power flow between the two focal spots.

For example, Mao *et al.* [22] explored a tuneable photonic nanojet formed by using the MGLL, which can achieve ultra-long subwavelength focusing between the two focal points. At microwave frequencies, Chou *et al.* [23] investigated the use of a metasurface comprising circular metallic patches printed on a grounded dielectric substrate to achieve the double-foci MGLL. Inspired by the optical double-foci GRIN MGLL [20], in this study, an acoustic double-foci GRIN MGLL is proposed for achieving a super long acoustic jet that covers both the near field and the far field. The variation of the refractive index of the lens is achieved by varying the filling ratio of lattice unit cells, which allows for tailoring the velocity of acoustic wave propagation in the structure. With this method, the graded change of the refractive index can be obtained in a broadband frequency range.

## 2. Double-foci Luneburg Lens Design

The double-foci MGLL design has two concentric circles (radii of $R$ and $R'$, $R$ is the radius of the lens) with different graded refractive index profiles. When a line source is used to generate plane waves that interact with the lens, the acoustic waves passing through the MGLL will produce double-foci, as shown in Figure 1 (a). All the rays entering the inner circle will be perfectly focused on the near field with a focal length of $F_1$ (blue dot), while the rays entering the outer circle will be focused to the far field with a focal length of $F_2$ (red dot). According to reference [20], the trajectories of rays can be described by the following set of equations:

$$\begin{cases} \int_r^{R'} \frac{k}{r\sqrt{\rho^2-k^2}} dr + \int_{R'}^{R} \frac{k}{r\sqrt{\rho^2-k^2}} dr = \frac{1}{2} \arcsin\left(\frac{k}{F_1}\right) + \arccos(k), & 0 \leq \rho \leq P_a \\ \int_r^{R} \frac{k}{r\sqrt{\rho^2-k^2}} dr = \frac{1}{2} \arcsin\left(\frac{k}{F_2}\right) + \arcsin(k), & P_a \leq \rho \leq 1 \end{cases} \quad (1)$$

where $r$ is the radial distance, $k = n(r)r \cdot \sin(\varphi)$, and $\varphi$ is the angle between the tangential direction and the position vector $r$. $P_a$ is a parameter between 0 and 1, which can be used to determine the radius of the inner circle of the double-foci Luneburg lens. $P_a$ related to the refractive index and the radius of the inner circle $R'$ can be described as $P_a = R'n(R')$, where $n(R')$ is the refractive index at the inner circle. $\rho = rn/R$, then the refractive index is obtained as:

$$n(\rho) = \begin{cases} e^{\{\omega_1(\rho,F_2)+\omega_2(\rho,F_1,P_a)-\omega_2(\rho,F_2,P_a)\}}, & 0 \leq \rho < P_a \\ e^{\omega_1(\rho,F_2)}, & P_a \leq \rho \leq 1 \end{cases} \quad (2)$$

where $\omega_1(\rho, F_2) = \frac{1}{\pi} \int_\rho^1 \frac{\arcsin(k/F_2)}{\sqrt{k^2-\rho^2}} dk$ and $\omega_2(\rho, F_i, P_a) = \frac{1}{\pi} \int_\rho^{P_a} \frac{\arcsin(k/F_i)}{\sqrt{k^2-\rho^2}} dk$.

Based on the distributions of the refractive index in Equation (2), it can be seen that the refractive index is a function of both near-field focal length $F_1$ and far-field focal length $F_2$. In this study, we fix the value of $F_1$ and tune the length of acoustic jet via changing the value of $F_2$. The refractive index distribution of the MGLL with respect to the radial distance $r$ at different values of $F_2$ (1.6R, 2.5R, and 3.2R) are provided in Figure 1(b).

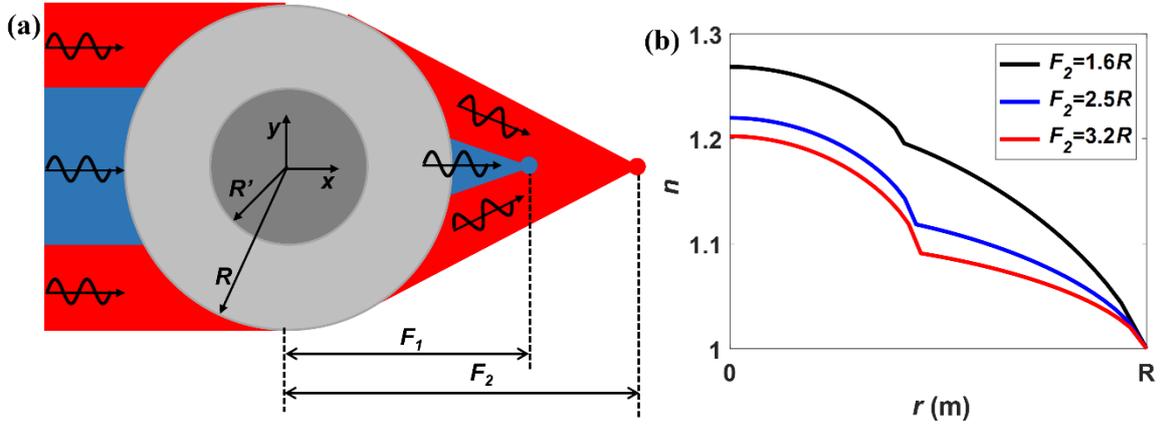

**Figure 1: Mechanism and design principle of the double-foci MGLL. (a) Schematic of the double-foci MGLL for manipulating acoustic waves. (b) The distribution of refractive index along the radial distance $r$ for different far field focal length $F_2$ (the near field focal length $F_1$ is fixed to be 1.2R).**

## 3. Impedance-Matching Double-foci Luneburg Lens

In this study, a perfect impedance-matching double-foci Luneburg lens is designed with continuous variation of refractive index according to Equation (2) to generate an acoustic jet. Numerical simulations were performed to investigate the acoustic jet performance. The lens design parameters were chosen to be $R=0.1$ m, $P_a=0.5$, $F_1=1.2R$, and different values of $F_2$ (1.6$R$, 2.5$R$, and 3.2$R$) were used. The dimension of the air area in the simulation was $7R \times 3R$. Frequency domain analysis was carried out by using the acoustic module of COMSOL software at the frequency $f$ = 17 kHz (wavelength $\lambda = 0.02$ m). A line source located at $x = -1.5R$ ($y$ is from -1.5$R$ to 1.5$R$) was used for excitation. In the simulations, radiation boundary conditions were applied on the air boundary to reduce the boundary reflections. Numerical simulation results are shown in Figure 2 (a)-(f) for different values of $F_2$, which clearly demonstrate that an ultra-long acoustic jet can be achieved. In addition, ray trajectories were calculated based on the ray tracing technique and overlayed with the simulated wave field in Figure 2(a)-(c), which exhibit good agreement with the numerical simulation results.

Furthermore, the acoustic intensity distributions of the double-foci MGLL are shown in Figure 2(d)-(f)) and the normalized intensity profiles along the $x$ and $y$ axes (along the white dash-dotted lines in Figure 2(d)-(f)) are plotted in Figure 2 (g) and (h). The jet length (JL) and full width at half maximum (FWHM) of the acoustic jet were obtained based on the acoustic intensity along $x$ and $y$ direction and shown in Figure 2(i) and (j). The JL is the length where the intensity is above the half of its maximum value along $x$ axis, which starts from the outer surface of the MGLL ($x$=0.1 m). The FWHM is the length between the two locations at which the intensity is equal to half of its maximum value along $y$ axis. It can be clearly seen that the obtained acoustic jet has a JL over 17$\lambda$ at $F_2$ =3.2$R$, which starts from the outer surface of the Luneburg lens (near field) and ends at a distance of more than 17$\lambda$ from the outer surface of the lens (far field). In addition, the length of the acoustic jet can be tuned by changing the focal length $F_2$.

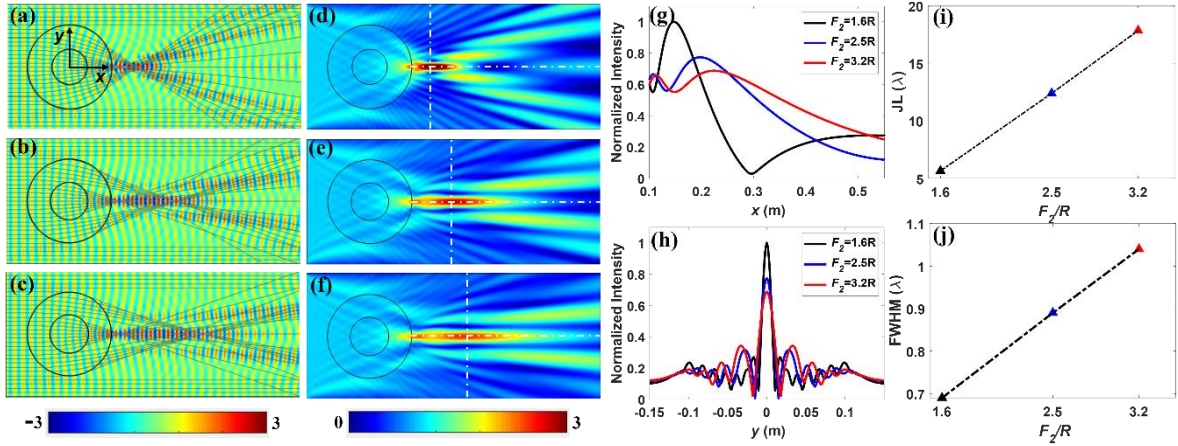

**Figure 2: Numerical simulations of steady state responses of perfect impedance matching double-foci MGLL. (a) - (c) Waveform fields for $F_2 = 1.6R$, $F_2 = 2.5R$, and $F_2 = 3.2R$, respectively, overlayed with the ray trajectories obtained with the ray tracing method. (d) – (f) Intensity distributions for $F_2 = 1.6R$, $F_2 = 2.5R$, and $F_2 = 3.2R$, respectively. (g) and (h) Normalized acoustic intensity along the *x* axis and the *y* axis for different focal length $F_2$ at the maximum intensity spots. (i) and (j) JL and FWHM obtained for different focal length $F_2$. The black circle indicates the outline of the Luneburg lens, and the colour indicates the acoustic field with a unit of Pa. The input acoustic pressure is 1 Pa.**

## 4. Acoustic Metamaterial Double-foci Luneburg Lens

Perfect impedance-matching MGLL is an ideal case, which can hardly be manufactured in real applications. In this study, GRIN acoustic metamaterials [24, 25] are used to realize a practical double-foci MGLL for obtaining acoustic jets. The acoustic metamaterial was designed to have the same geometric parameters as the impedance-matching double-foci Luneburg lens: $R$=0.1 m, $P_a$=0.5, $F_1$=1.2$R$, and three different values of $F_2$ (1.6$R$, 2.5$R$, and 3.2$R$). To obtain the refractive index distributions shown in Figure 1(b), various unit cells can be used. Here, we used a unit cell of three-dimensional (3D) lattice [26, 27], as shown in Figure 3(a), which has three orthogonal beams, so that each unit cell is interconnected with its adjacent cells to form a self-supported lattice. The unit cell used to build the 2D lens is a 3D truss unit cell. The Luneburg lens made of such a 3D unit cell renders the lens to be stiff and free of deformation, while being lightweight.

In the following studies, the periodicity of the unit cell $D$ is chosen to be 5 mm. In order to determine the relation between the refractive index with the lattice unit cell, dispersion curves along the $\Gamma X$ direction based on the primitive cubic (as shown in Figure 3(b)) was calculated by using Comsol software. The filling ratio of the unit cell can be changed through tailoring the factor $a_0$, and hence the dispersion curve can be changed correspondingly (see Figure 3(c)). The refractive index of each unit cell was calculated based on the slope of the dispersion curve [28]. Note that the slope of the dispersion curve is almost a constant value in the broadband frequency range of 0 to 20 kHz, which indicates that the refractive index is independent of frequency, and hence the acoustic jet can be achieved in a broadband frequency range. Note that based on the homogeneity condition [29], the working frequency is limited by the geometric parameters of the lens as $4D<\lambda<R$. In this work, we designed two-dimensional (2D) double-foci Luneburg lens working at the frequency range $f = 11$ kHz – 17 kHz with three different $F_2$ values (1.6$R$, 2.5$R$, and 3.2$R$), an example of $F_2 = 1.6R$ is shown in the inset of Figure 3(c).

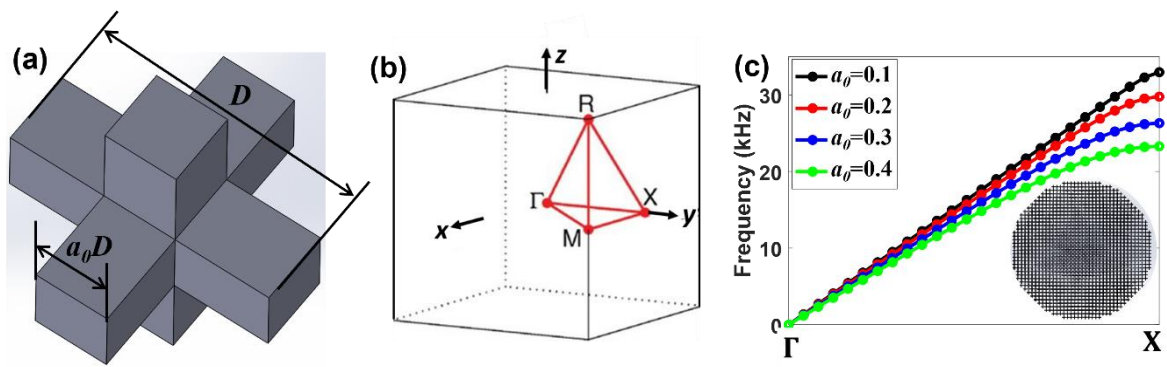

**Figure 3: Acoustic metamaterial double-foci MGLL design. (a) An example of 3D lattice unit cell. (b) Primitive cubic unit cell used for dispersion curves. (c) Dispersion curves for**

**the unit cell with different parameter *a₀* with an inset of acoustic metamaterial double-foci Luneburg lenses having *F₂* = 1.6*R*.**

In this study, we explored the capability of the acoustic metamaterial double-foci Luneburg lens for producing an acoustic jet. Full 3D wave simulations were conducted using commercial Comsol software. Similar as in the impedance-matching Luneburg lens simulations, radiation boundary conditions were applied on the outer boundaries of the lens to assume infinite air spaces. The dimension of the air area in the simulation was $7R \times 4R$. A plane wave of 17 kHz was used for excitation at the location of $x = -1.2R$ ($y$ from $-2R$ to $2R$). The numerical simulation results for both the waveform fields and intensity distributions are shown in Figure 4 (a)-(f). The acoustic metamaterial MGLL is demonstrated to have the capability of generating an ultra-long acoustic jet, similar to that of the perfect impedance-matching Luneburg lens.

Similarly, the acoustic intensity profiles of the acoustic metamaterial double-foci MGLL along the *x* and *y* axes of the lens at the maximum intensity spots along the white dash-dotted lines in Figure 4(d)-(f) are plotted and shown in Figure 4 (g) and (h). The JL and the FWMH were obtained and shown in Figure 4(i) and (j). Again, the acoustic metamaterial MGLL is proven to be able to achieve an ultra-long acoustic jet starting from the near field (outer surface of the lens) and extending to the far field (with JL is more than $13\lambda$ for $F_2 =3.2R$), and the FWHMs are around $\lambda$.

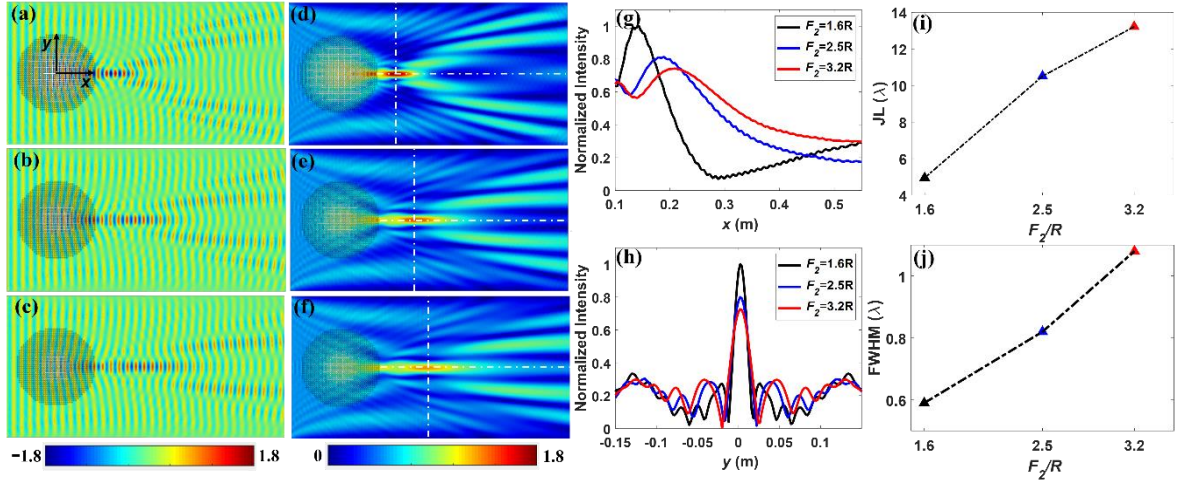

**Figure 4: Numerical simulation results of the acoustic metamaterial double-foci Luneburg lens with different focal length $F_2$ for generating acoustic jet at 17 kHz. (a) - (c) Waveform fields obtained for $F_2 = 1.6R$, $F_2 = 2.5R$, and $F_2 = 3.2R$, respectively. (d) − (f) Intensity distributions obtained for $F_2 = 1.6R$, $F_2 = 2.5R$, and $F_2 = 3.2R$, respectively. (g) and (h) Intensity profiles along the *x* axis and the *y* axis at the maximum intensity spots. (i) and (j) JL and FWHM obtained for different focal length $F_2$. The colour indicates the acoustic field with a unit of Pa. The input acoustic pressure is 1 Pa.**

In order to investigate the broadband characteristic of the acoustic metamaterial MGLL, excitation signals of different frequencies ($f$ = 11 kHz, 13 kHz, and 15 kHz) were used for the MGLL with a focal length of $F_2$ =2.5$R$. The simulated waveform fields and intensity distributions at different frequencies are shown in Figure S1. In addition, the acoustic intensity profiles of the acoustic metamaterial double-foci MGLL along the *x* direction and the corresponding JL were obtained and shown in Figure S2. These results reveal the excellent performance of the MGLL for producing ultra-long acoustic jets in a broadband frequency range. The JL is higher at a lower frequency and a maximum JL of over 30$\lambda$ can be obtained at 11 kHz. Note that the upper frequency range of the acoustic metamaterial double-foci MGLL scales linearly with the unit cell size without the change of JL/λ. Therefore, the lens can also be designed to work at higher frequencies by using smaller unit cell dimensions.

In addition, a quantitative comparison of perfect impedance-matching MGLL and acoustic metamaterial MGLL was performed. The error in the jet length between the perfect impedance-matching MGLL and acoustic metamaterial MGLL is provided in Figure S3, which is calculated according to $Error = \frac{|JL(PL) - JL(AM)|}{JL(PL)}$, where $JL(PL)$ is the jet length of the perfect impedance-matching MGLL and $JL(AM)$ is the jet length of the acoustic metamaterial MGLL. This error is due to the discrete refractive indices of the acoustic metamaterial MGLL, which generates impedance mismatch between unit cells. As a result, part of the energy is reflected back and cannot propagate through the lens. For example, based on the simulations, there is almost no reflected energy for the perfect impedance-matching MGLL, while for the acoustic metamaterial MGLL with $F_2 = 2.5R$, around 15% of the energy is reflected back at the frequency of 17 kHz.

## 5. Conclusions

In this paper, we proposed a novel design of gradient index (GRIN) acoustic metamaterial lens based on the modified generalized Luneburg lens (MGLL) to realize double foci, which can help achieve an ultra-long acoustic jet in a broadband frequency range. Theoretical studies and numerical simulations were carried out to investigate the capability of the proposed lens for generating ultra-long acoustic jet. The acoustic metamaterial MGLL was demonstrated to have a JL up to 30$\lambda$ covering both the near field and the far field.

**Conflict of Interest**

The authors declare no conflict of interest.

**Acknowledgements**

This work was supported by the AFOSR Center of Excellence on Nature-Inspired Flight Technologies and Ideas and NSF (CMMI 1436347).

**Dada Availability**

The data that support the findings of this study are available from the corresponding authors upon reasonable request.